\documentclass[a4paper,10pt,intlimits,twocolumn,aps,pre,superscriptaddress,nofootinbib]{revtex4-1}
\usepackage{amssymb,amsfonts,amsmath,amsthm,mathtools,dsfont}
\usepackage[justification=justified,singlelinecheck=false]{caption,subfig}
\usepackage{tikz,pgfplotstable}
\usetikzlibrary{positioning,shadows,backgrounds,shapes.geometric}
\usepackage{enumerate,stmaryrd,color}

\makeatletter
\newsavebox{\mybox}\newsavebox{\mysim}
\newcommand{\asym}[1]{%
  \savebox{\mybox}{\hbox{\kern3pt$\scriptstyle#1$\kern3pt}}%
  \savebox{\mysim}{\hbox{$\sim$}}%
  \mathbin{\overset{#1}{\kern\z@\resizebox{\wd\mybox}{\ht\mysim}{$\sim$}}}%
}

\makeatother

\newcommand{\beq}{\begin{equation}}
\newcommand{\eeq}{\end{equation}}
\newcommand{\bea}{\begin{eqnarray}}
\newcommand{\eea}{\end{eqnarray}}

\theoremstyle{definition}
\newtheorem{dalpha}{Definition}[section]
\newtheorem{dfac}[dalpha]{Definition}
\newtheorem{dbino}[dalpha]{Definition}
\newtheorem{dbeta}[dalpha]{Definition}

\theoremstyle{plain}
\newtheorem{thalpha}{Theorem}[section]
\newtheorem{thbeta}[thalpha]{Theorem}
\newtheorem*{thalpha2}{Theorem}
\newtheorem*{thbeta2}{Theorem}

\begin{document}
\title{On the robustness of the $q$--Gaussian family}
\author{Gabriele Sicuro}\email{sicuro@cbpf.com}
\affiliation{Centro Brasileiro de Pesquisas F\'isicas, Rua Dr. Xavier Sigaud, 150, 22290-180, Rio de Janeiro, Brazil}
\author{Piergiulio Tempesta}
\affiliation{Departamento de F\'{\i}sica Te\'{o}rica II (M\'{e}todos Matem\'{a}ticos de la f\'isica), Facultad de F\'{\i}sicas, Universidad
Complutense de Madrid, 28040, Madrid, Spain, and
Instituto de Ciencias Matem\'aticas, C/ Nicol\'as Cabrera, No 13--15, 28049, Madrid, Spain}
\author{Antonio Rodr\'iguez}
\affiliation{GISC, Departamento de Matem\'atica Aplicada a la Ingenier\'ia Aeroespacial, Universidad Polit\'ecnica de Madrid, Plaza Cardenal Cisneros s/n, 28040, Madrid, Spain}
\author{Constantino Tsallis}
\address{Centro Brasileiro de Pesquisas F\'isicas, Rua Dr. Xavier Sigaud, 150, 22290-180, Rio de Janeiro, Brazil}
\affiliation{National Institute of Science and Technology for Complex Systems,  Rua Dr. Xavier Sigaud, 150, 22290-180, Rio de Janeiro, Brazil, and Santa Fe Institute, 1399 Hyde Park Road, Santa Fe, New Mexico, 87501 USA}

\begin{abstract}
We introduce three deformations, called $\alpha$--, $\beta$-- and $\gamma$--deformation respectively, of a $N$--body probabilistic model, first proposed by Rodr\'iguez et al.\@ (2008), having $q$--Gaussians as $N\to\infty$ limiting probability distributions. The proposed $\alpha$-- and $\beta$--deformations are asymptotically scale--invariant, whereas the $\gamma$--deformation is not. We prove that, for both $\alpha$-- and $\beta$--deformations, the resulting deformed triangles still have $q$--Gaussians as limiting distributions, with a value of $q$ independent (dependent) on the deformation parameter in the $\alpha$--case ($\beta$--case). In contrast, the $\gamma$--case, where we have used the celebrated $Q$--numbers and the Gauss binomial coefficients, yields other limiting probability distribution functions, outside the $q$--Gaussian family. These results suggest that scale--invariance might play an important role regarding the robustness of the $q$--Gaussian family. 
\end{abstract}
\date{\today}
\maketitle
\section{Introduction}
It is well known that the (properly centered and rescaled) sum of $N$ independent (or weakly dependent) random variables with finite variance approaches a Gaussian distribution in the $N\to\infty$ limit \cite{feller1971}. This fundamental classical result, known as the Central Limit Theorem (CLT), is at the basis of the ubiquity of Gaussian distributions in Nature. The classical CLT, however, cannot be applied to a set of strongly correlated random variables. Therefore, in the context of nonextensive statistical mechanics \cite{tsallis2009}, it has been argued the existence of a generalized CLT for random variables correlated in a specific way \cite{umarov2008,umarov2010}, called $q$--independence. Alternative CLTs, based on a different kind of correlations \cite{vignat2007} or on exchangeability \cite{hahn2010}, have also been proposed in the literature. In all these theorems, the (properly centered and rescaled) sum of $N$ correlated random variable has, in the $N\to\infty$ limit, a \textit{$q$--Gaussian distribution} $P_q(x)$,
\begin{equation}
P_q(x)\coloneqq\begin{cases}\sqrt{\frac{q-1}{\pi}}\frac{\Gamma\left(\frac{1}{q-1}\right)}{\Gamma\left(\frac{3-q}{2q-2}\right)}e_q^{-x^2}&q\in(1,3),\\ \frac{1}{\sqrt{\pi}}e^{-x^2}&q=1,\\ \sqrt{\frac{1-q}{\pi}}\frac{3-q}{2}\frac{\Gamma\left(\frac{3-q}{2-2q}\right)}{\Gamma\left(\frac{1}{1-q}\right)}e_q^{-x^2}&q\in(-\infty,1),   \end{cases}
\end{equation}
where $q$ is a real parameter depending on the nature of the correlations, and the \textit{$q$--exponential function} is defined as follows:
\begin{equation}
e_q^{x}\coloneqq\begin{cases}e^x&q=1,\\\left[1+(1-q)x\right]_+^\frac{1}{1-q}&q\neq 1,\end{cases}
\end{equation}
with $[x]_+\coloneqq x\, \theta(x)$, $\theta(x)$ Heaviside function. In what follows, we will use also the inverse function of the $q$--exponential, the \textit{$q$--logarithm}, defined as
\begin{equation}
\log_q(x)\coloneqq\begin{cases}\ln(x)&q=1,\\\frac{x^{1-q}-1}{1-q}&q\neq 1,\end{cases}\qquad x>0.
\end{equation}

Like Gaussian distributions, $q$--Gaussians also are ubiquitous in Nature. Indeed, analytical, experimental and numerical investigations in biology \cite{upadhyaya2001}, economics \cite{tsallis2003,borland2002}, high energy physics \cite{wong2013}, anomalous diffusion processes \cite{tsallis1996anomalous,andrade2010}, dynamics of many-body classical Hamiltonian systems \cite{anteneodo1998,leo2012non,cirto2014,pluchino2007}, cold atoms \cite{douglas2006tunable,lutz2003anomalous,lutz2013beyond}, dissipative and conservative low dimensional maps \cite{tirnakli2007central,tirnakli2015standard}, turbulence \cite{boghosian1996thermodynamic} among others\footnote{For a regularly updated bibliography on nonextensive thermostatistics and related topics, see \texttt{http://tsallis.cat.cbpf.br/biblio.htm}}, have shown that $q$--Gaussian distributions appear in the probabilistic analysis of many systems in which long--range interactions are present, or ergodicity lacks. These evidences strongly support the existence of a generalized CLT involving $q$--Gaussians.

To investigate the conditions under which such a generalized CLT holds, {\textit{analytically solvable} probabilistic models yielding $q$--Gaussian limiting distributions are of paramount importance}. In particular, in \cite{rodriguez2008} a probabilistic model {for $N$ correlated binary random variables} was introduced, generalizing the celebrated \textit{Leibniz triangle} \cite{polya1962}. {%The Leibniz triangle is a probabilistic model describing a set of $N$ identical correlated binary random variables $\{x_i\}_{i=1,\dots,N}$, $x_i\in\{0,1\}$, such that the quantity $n\coloneqq\sum_{i=1}^Nx_i$ is uniformly distributed between $0$ and $N$ for $N\to\infty$: the model in \cite{rodriguez2008} is obtained from the Leibniz triangle introducing an additional constraint on the admissible values of $n$. 
The proposed model preserves the {\em scale--invariance property} (see Ref.~\cite{rodriguez2008} and below for a definition) of the Leibniz triangle and it can be rigorously proved that this model has $q$--Gaussians with $q\in[0,1]$ as limiting distributions for $N\to\infty$}. %An interpretation of this long--range correlated model as a nonlinear coupling of statistical states has been recently proposed by \citet{nelson2010}.
Subsequently, it was shown \cite{rodriguez2012,rodriguez2014} that particular $d$--dimensional scale--invariant probabilistic models with $d\geq 1$ have, as limiting distributions, Dirichlet distributions for $d>1$, whereas for $d=1$ $q$--Gaussians were obtained. The ultimate relationship between scale--invariance and $q$--{\em Gaussianity}, i.e.\@ the appearance of $q$--Gaussians as probability distributions for statistical models in the thermodynamic limit, is not yet completely clear. In Refs.~\cite{rodriguez2008,tsallis2005} it was already conjectured that (asymptotic) scale invariance could be possibly a necessary but not sufficient condition, for the emergence of $q$-Gaussians. {Indeed, in \cite{hilhorst2007} two scale--invariant probabilistic models which are different from the one analyzed here, are analytically studied, showing a limiting distribution different from a $q$--Gaussian}.

With the aim of shedding further light on this problem, we address here the \textit{robustness} under small perturbations of the general one--parameter family of scale--invariant probabilistic models introduced in \cite{rodriguez2008}. {In particular, we investigate the stability of the $q$--Gaussian family in the space of probability distributions: this property is indeed fundamental for the existence of a generalized CLT yielding $q$--Gaussian distributions in Nature. In this context, exactly solvable probabilistic models are essential tools for a rigorous study of the type of correlations and the properties required for such generalized theorem}. 

We consider two new families of {\em asymptotically scale--invariant} triangles, namely the $\alpha$--{\em triangles} and $\beta$--{\em triangles}, which generalize the aforementioned family. These deformations are based on the introduction of two classes of real numbers, the \textit{$\alpha$--numbers} and \textit{$\beta$--numbers} respectively, in the same spirit of the \textit{$Q$--numbers}, typical of the $Q$--deformations of Lie groups and algebras \cite{kac2002}. As we shall see, despite the deformation, the limiting distributions remain $q$--Gaussians, but with a value of $q$ which might depend on the perturbation strength.

To the best of our knowledge, this is the first article addressing, for specific probabilistic models, the important problem of the robustness of $q$--Gaussians as \textit{attractors}, a fundamental property involved in the existence of a generalized CLT.

The aforementioned deformations may be considered as nontrivial ones, since there is no a priori guarantee that an arbitrary deformation should preserve the same $q$--Gaussian behavior for large values of $N$. As a counterexample, we introduce and study here an alternative deformation, that we call \textit{$\gamma$--deformation}, based on the classical definition of $Q$--number \cite{kac2002} used in combinatorics. This deformation does not generically preserve $q$--Gaussian forms for the limiting distributions. Since scale--invariance is violated by the $\gamma$--deformation, in contrast with the $\alpha$-- and $\beta$-- ones, a possible conjecture might emerge on the necessity of (at least asymptotic) scale-invariance for $q$--Gaussianity (see also \cite{tsallis2005}).

%The paper is organized as follows. In Section 2, the mathematical apparatus necessary for the subsequent %discussion is proposed. In Section 3, a family of statistical models, depending on a real parameter $\alpha$ is %introduced, and further generalized in Section 4. In Section 5 we present a theorem proving the robustness of %our models. In Section 6, we introduce other models, that are not robust, in order to show the nontriviality of %the deformations previously discussed. Future perspectives and open problems are discussed in the final Section %7.

\section{Preliminaries: Leibniz-like triangles as probability models}\label{LT}
Let us consider a system with $N$ \textit{identical} elements, whose states are  characterized by  \textit{binary} variables $x_i\in\{0,1\}$, $i=1,\,\dots,N$. Let us introduce also the probability $r_{N,n}$ of having a given configuration $\{x_1,\dots,x_N\}$ with $\sum_{i=1}^Nx_i=n$. The probability $p_{N,n}$ of having any configuration such that $\sum_{i=1}^Nx_i=n$ is obtained taking into account the proper degeneracy
\begin{equation}
p_{N,n}\coloneqq\binom{N}{n}r_{N,n}.
\end{equation}
The set of values $r_{N,n}$ can be organized in a triangle, in such a way that the element $r_{N,n}$ is the $n$th element of the $N$th row. We require that the following \textit{Leibniz triangle rule} (or \textit{scale--invariance property}) holds:
\begin{equation}\label{scale_invariance_condition}
r_{N,n+1}+r_{N,n}=r_{N-1,n}.
\end{equation}
The \textit{Leibniz triangle} $r^{(1)}_{N,n}$ \cite{polya1962} can be constructed using the Leibniz rule and defining $r_{N,0}^{(1)}$ as follows:
\begin{equation}r^{(1)}_{N,0}\coloneqq \frac{1}{N+1}\Rightarrow
r^{(1)}_{N,n}=\frac{1}{\binom{N}{n}}\frac{1}{N+1}.\label{Leibniz_triangle}
\end{equation}
The $N\to\infty$ limiting distribution is the \textit{uniform distribution}. Considering instead $r_{N,0}=p^N$, $p\in(0,1)$, the $N\to\infty$ limiting distribution is a \textit{Gaussian distribution}, being in this case $p_{N,n}=\binom{N}{n}p^n(1-p)^{N-n}$, i.e.,  the binomial distribution. 

The aforementioned Leibniz triangle has been generalized in \cite{rodriguez2008}: for $\nu\in\mathds N$,
\begin{equation}
r^{(\nu)}_{N,0}\coloneqq \frac{\Gamma(2\nu)\Gamma(N+\nu)}{\Gamma(\nu)\Gamma(N+2\nu)}\Rightarrow r^{(\nu)}_{N,n}=\frac{r^{(1)}_{N+2(\nu-1),n+\nu-1}}{r^{(1)}_{2(\nu-1),\nu-1}}.\label{r_nu}
\end{equation}
The triangle \eqref{Leibniz_triangle} is recovered as the $\nu =1$ particular case.
Remarkably, it has been proved \cite{rodriguez2008} that, for $N\to\infty$ and $\nu=2,3,\dots$, we have\footnote{Observe that in $\sum_{k=0}^N\binom{N}{k}r^{(\nu)}_{N,k}=1$; this result can be proved using the identity $\frac{1}{n+1}\binom{n}{k}^{-1}=\int_0^1x^k(1-x)^{n-k}dx$ and the binomial theorem.}
\begin{multline}
p^{(\nu)}_{N,n}\coloneqq\binom{N}{n}r^{(\nu)}_{N,n}\asym{N\gg 1}\frac{2\sqrt{\nu-1}}{N}P_{q_1(\nu)}(x),\\ x\coloneqq2\sqrt{\nu-1}\left(\frac{n}{N}-\frac{1}{2}\right),
\end{multline}
where $P_{q_1(\nu)}(x)$ is a $q$--Gaussian with $q\equiv q_1(\nu)$,
\begin{equation}
q_1(\nu)=1-\frac{1}{\nu-1}<1\,;
\end{equation}
the subindex 1 will become transparent later on.
In the $\nu\to\infty$ limit, $r^{(\nu)}_{N,0}\to 2^{-N}$ and $q_1(\nu)\to 1$ as expected from the CLT.

\section{Asymptotically scale--invariant deformations of the generalized Leibniz triangle} \label{beta}
In this Section, we introduce two asymptotically scale--invariant deformations of the probabilistic model analyzed in \cite{rodriguez2008}.

\subsection{The $\alpha$--numbers and the $\alpha$--triangles}
The basis of our construction is the notion of $\alpha$--number.
\begin{dalpha}
Given $n\in\mathds{N}\cup{0}$, and $\alpha>0$, $\alpha\neq 1$, an {\rm$\alpha$--number} is the real number defined as follows:
\begin{equation}
\{n\}_\alpha\coloneqq (n+1)\left(1-\frac{1-\alpha}{1-{\alpha}^{n+1}}\right)=\begin{cases}n a_n(\alpha)&n\geq 1\,,\\ 0&n=0 \,,\end{cases}\label{alpha_number}
\end{equation}
where we have introduced ($n\geq 1$)
\begin{equation}
a_n(\alpha)\coloneqq\frac{n+1}{n}\left(1-\frac{1-\alpha}{1-{\alpha}^{n+1}}\right)\xrightarrow{\alpha\to 1}1\,.
\end{equation}
\end{dalpha}
The previous definition is such that
\begin{equation}\{n\}_1\equiv\lim_{\alpha\to 1}\{n\}_\alpha=n \,,\label{alpha_to_1_limit}\end{equation}
so in the $\alpha\to 1$ limit we recover the ordinary numbers.

The following is a generalization of the factorial of a natural number.

\begin{dfac}
Given $n\in\mathds{N} \cup 0$, we shall call {\it$\alpha$--factorial} the number defined as
\begin{equation}
\{0\}_\alpha!\coloneqq 1\text{ and }\{n\}_\alpha!\coloneqq\prod_{k=1}^{n}\{k\}_\alpha=n!\prod_{k=1}^{n}a_{k}(\alpha),\ n\in\mathds N.
\end{equation}
\end{dfac}

The ordinary factorial is recovered in the $\alpha\to 1$ limit, $\{n\}_1!\equiv\lim_{\alpha\to 1}\{n\}_\alpha!=n!$. We define now an extension of the binomial coefficients.

\begin{dbino}
Given the nonnegative integers $N$ and $n\leq N$, the {\it $\alpha$--binomial coefficient} is defined as
\begin{multline}
\left\{\begin{matrix}N\\n\end{matrix}\right\}_\alpha\coloneqq\frac{\{N\}_\alpha!}{\{n\}_\alpha!\{N-n\}_\alpha!}\\
=\begin{cases}1&\text{if $n=0$ or $n=N$},\\\binom{N}{n}\prod_{k=1}^{N-n}\frac{a_{k+n}(\alpha)}{a_{k}(\alpha)}&1\leq n \leq N-1.\end{cases}
\end{multline}
\end{dbino}
The $\alpha$--binomial coefficients share with the Pascal coefficients the property $\left\{\begin{smallmatrix}N\\n\end{smallmatrix}\right\}_\alpha=\left\{\begin{smallmatrix}N\\N-n\end{smallmatrix}\right\}_\alpha$, $\forall\alpha$. Again, we recover the ordinary binomial coefficients in the $\alpha\to 1$ limit. In the same fashion as the Pascal triangle, the $\alpha$--binomial coefficients can be displayed forming a {\em Pascal $\alpha$--triangle}:
\begin{center}
\includegraphics[scale=1]{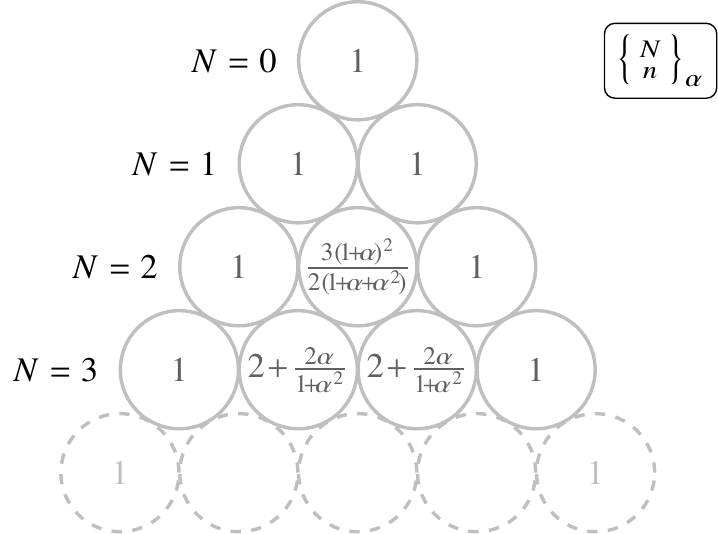}
\end{center}

\paragraph{Deformation of the Leibniz--like triangles using the $\alpha$--numbers}

We want now to deform the family of triangles obtained in \cite{rodriguez2008} using the aforementioned $\alpha$--numbers. Let us start introducing the {\em Leibniz--like $\alpha$--triangle} as
\begin{equation}
r^{(1)}_{N,n,\alpha}\coloneqq \frac{1}{\{N+1\}_\alpha\left\{\begin{smallmatrix}N\\n\end{smallmatrix}\right\}_{\alpha}},\quad  n=0,\dots,N,\label{alpha-Leibniz}
\end{equation}
which is related to the original Leibniz triangle as
\begin{equation}
r^{(1)}_{N,n,\alpha}=\mu^{(1)}_{N,n,\alpha}r^{(1)}_{N,n},
\label{relacion_alpha}\end{equation}
with
\begin{multline}
\mu^{(1)}_{N,n,\alpha}\coloneqq\\\begin{cases}1&\text{for $N=0$,}\\\dfrac{1}{a_{N+1}(\alpha)}&\text{for $N>0$ and $n=0$ or $n=N$,}\\\displaystyle\frac{\prod_{k=1}^{N-n}\frac{a_k(\alpha)}{a_{k+n}(\alpha)}}{a_{N+1}(\alpha)}&\text{for $N>0$ and $n=1,\dots,N-1$}.
\end{cases}
\end{multline}
In addition, $\lim_{\alpha\to 1}\mu^{(1)}_{N,n,\alpha}=1$. As an illustration, the $\alpha=\frac{1}{2}$ instance of family \eqref{alpha-Leibniz} is given by
\begin{center}
\includegraphics[scale=1]{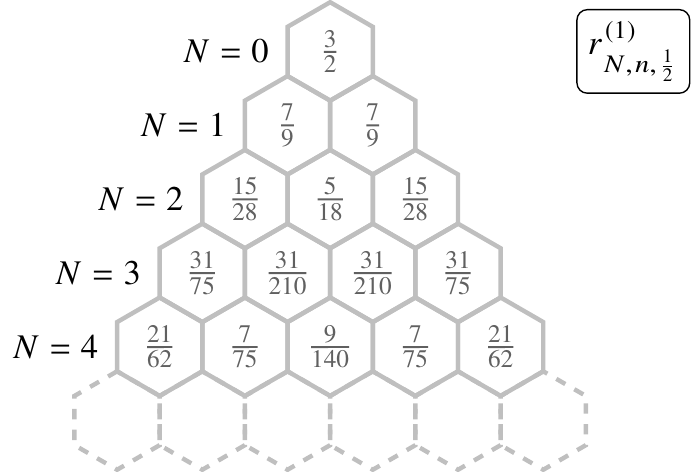}
\end{center}

The product of the Leibniz $\alpha$--triangle by the Pascal triangle (which takes into account the degeneracies) does {\em not} generically yield a set of probabilities, since \[\sum_{n=0}^N\binom{N}{n}r^{(1)}_{N,n,\alpha}\neq 1.\] Nevertheless we can circumvent this difficulty by properly renormalizing the triangle to get a new one with coefficients
\begin{equation}
\hat r^{(1)}_{N,n,\alpha}\coloneqq\dfrac{r^{(1)}_{N,n,\alpha}}{\sum_{k=0}^N\binom{N}{k}r^{(1)}_{N,k,\alpha}},
\label{triangulo_normalizado}
\end{equation}
and associated probabilities
\begin{equation}
\hat p^{(1)}_{N,n,\alpha}\coloneqq\binom{N}{n}\hat r^{(1)}_{N,n,\alpha}.
\label{probabilidad_normalizada_alpha}\end{equation}

The normalized Leibniz $\alpha$--triangle for $\alpha=\frac{1}{2}$ looks like
\begin{center}
\includegraphics[scale=1]{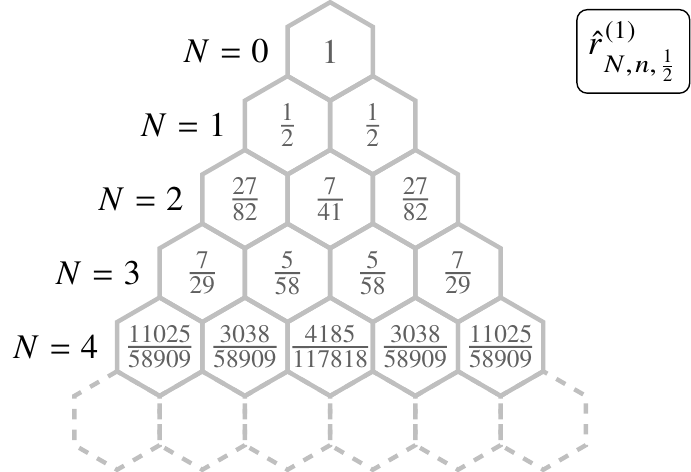}
\end{center}

Following Ref.~\cite{rodriguez2008}, we can define a two--parametric family of triangles from the Leibniz $\alpha$--triangle \eqref{alpha-Leibniz} for $\alpha>0$ and $\nu\in\mathds N$:
\begin{equation}r^{(\nu)}_{N,n,\alpha}\coloneqq\frac{r^{(1)}_{N+2(\nu-1),n+\nu-1,\alpha}}{r^{(1)}_{2(\nu-1),\nu-1,\alpha}}=\mu^{(\nu)}_{N,n,\alpha}r^{(\nu)}_{N,n},
\label{r_nu_alpha}\end{equation}
where, for $\nu>1$, we have\footnote{To obtain Eq.~\eqref{MuAlpha} we used the following identities for $K\in\mathds N$ and $a,b\in\mathds R^+$:
\begin{equation}\textstyle
\prod_{k=1}^K\frac{k+a+1}{k+a}=\frac{1+a+K}{1+a},\quad \prod_{k=1}^K\left(1-\frac{1-a}{1-a^{k+b}}\right)=\frac{a^{b+K}-a^{K}}{a^{b+K}-1}.
\end{equation}}
\begin{multline}\label{MuAlpha}
\mu^{(\nu)}_{N,n,\alpha}\coloneqq\\
\frac{a_{2\nu-1}(\alpha)}{a_{N+2\nu-1}(\alpha)}\prod_{k=1}^{N-n+\nu-1}\frac{a_k(\alpha)}{a_{k+n+\nu-1}(\alpha)}\prod_{k=1}^{\nu-1}\frac{a_{k+\nu-1}(\alpha)}{a_k(\alpha)}\\
=\frac{a_{2\nu-1}(\alpha)(\alpha-1)\left(\alpha^{N+2\nu}-1\right)}{\alpha(\alpha^{N-n+\nu}-1)(\alpha^{n+\nu}-1)}\cdot\\
\cdot \frac{(N-n+\nu)(n+\nu)}{N+2\nu}\prod_{k=1}^{\nu-1}\frac{a_{k+\nu-1}(\alpha)}{a_k(\alpha)}.
\end{multline}

As before, normalization is needed to obtain the family
\begin{equation}
\hat r_{N,n,\alpha}^{(\nu)}\coloneqq\frac{r_{N,n,\alpha}^{(\nu)}}{\sum_{k=0}^N\binom{N}{k}r_{N,k,\alpha}^{(\nu)}},\label{r_nu_alpha_normalizado}
\end{equation}
with associated probabilities
\begin{equation}
\hat p_{N,n,\alpha}^{(\nu)}\coloneqq\binom{N}{n}\hat r_{N,n,\alpha}^{(\nu)}\label{P_nu_alpha_normalizado}
\end{equation}
trivially satisfying $\sum_{n=0}^N\hat p_{N,n,\alpha}^{(\nu)}=1$. 

Observe that triangle \eqref{r_nu_alpha_normalizado} does {\em not} generically fulfill the scale--invariance condition \eqref{scale_invariance_condition}. Nevertheless, it is {\em asymptotically scale--invariant}, i.e.,
\begin{equation}
\lim_{\substack{N\to\infty\\\frac{n}{N}\equiv\eta\text{ fixed}}}\frac{\hat r^{(\nu)}_{N-1,n, \alpha}}{\hat r^{(\nu)}_{N,n, \alpha}+\hat r^{(\nu)}_{N,n+1, \alpha}}=1.\label{asymptotic_scale_invariant}
\end{equation}
This follows from the fact that the normalization $\sum_{n=0}^N\binom{N}{n}r_{N,n,\alpha}^{(\nu)}$ has a power--law scaling behavior (as we will see later by evaluating $\binom{N}{n}r^{(\nu)}_{N,n,\beta}$ for large $N$), and from the limit
\begin{equation}
\frac{r^{(1)}_{N-1+2(\nu-1),n+\nu-1, \alpha}}{r^{(1)}_{N+2(\nu-1),n+\nu-1, \alpha}+ r^{(1)}_{N+2(\nu-1),n+\nu, \alpha}}
\xrightarrow{N\to\infty}1,
\end{equation}
obtained using the limit $a_N\xrightarrow{N\to\infty}\min\{1,\alpha\}$.

\paragraph{A theorem on the robustness of Leibniz--like $\alpha$--triangles}
\begin{figure*}\centering
\includegraphics[scale=1]{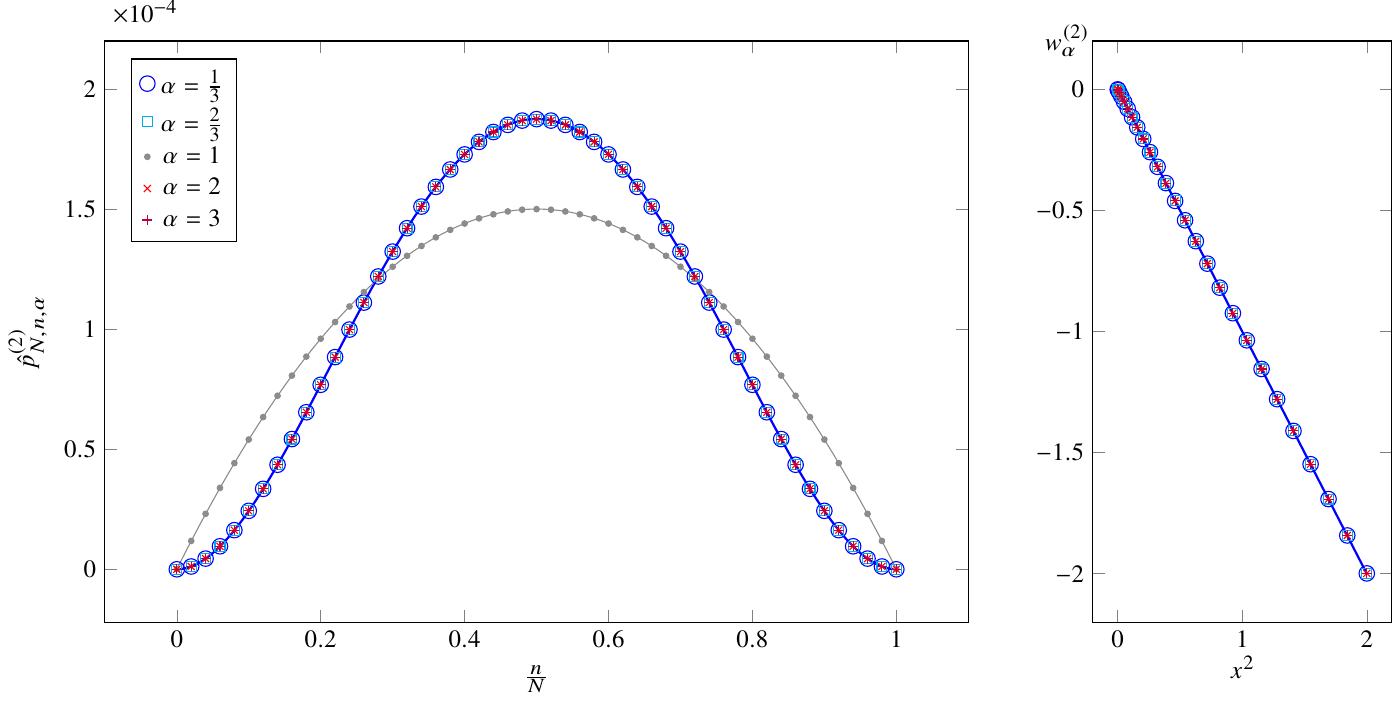}
\caption{Probability distributions \eqref{P_nu_alpha_normalizado} and $w^{(\nu)}_{\alpha}(x)\coloneqq\log_{q_\alpha(\nu)}\frac{P_{q_\alpha(\nu)}(x)}{P_{q_\alpha(\nu)}(0)}$ for $\nu=2$, $N=10^4$ and different values of $\alpha$; the derived asymptotic behavior is also depicted as continuous line for $\alpha\neq 1$.\label{fPa} The corresponding values of $q_\alpha(2)$ are given by Eq.~\eqref{qalpha}.
}
\end{figure*}

In this Section, we prove that the family of deformed triangles \eqref{r_nu_alpha_normalizado} still possesses a $q$--Gaussian as limiting distribution for $N\rightarrow \infty$, after proper centering and rescaling. However, the limiting value of $q$ is different from $q_1(\nu)$, {\it although not dependent on $\alpha$}.
\begin{thalpha}\label{ThAlpha}
The family of triangles \eqref{r_nu_alpha} with $\nu\in(1,+\infty)$ and $\alpha\in\mathds R^+\setminus\{0\}$ satisfies the property
\begin{equation}\label{EqThA}
\frac{N}{2\sqrt{\nu-\delta_{\alpha,1}}}\hat p_{N,n,\alpha}^{(\nu)} \asym{N\gg 1} P_{q_\alpha(\nu)}(x),
\end{equation}
where we have introduced the properly centered and rescaled variable
\begin{equation}\label{xvar}
x\coloneqq 2\sqrt{\nu-\delta_{\alpha,1}}\left(\frac{n}{N}-\frac{1}{2}\right).
\end{equation}
In Eq.~\eqref{EqThA}, $P_{q_\alpha(\nu)}(x)$ is a $q$--Gaussian with 
\begin{equation}\label{qalpha}
q_\alpha(\nu)\coloneqq 1-\frac{1}{\nu-\delta_{\alpha,1}}=\begin{cases}1-\frac{1}{\nu}&\text{for $\alpha\neq 1$},\\1-\frac{1}{\nu-1}&\text{for $\alpha=1$.} \end{cases}\end{equation}
\end{thalpha}

%\begin{proof} 
See Appendix \ref{A-alpha} for the proof. \\
%\end{proof}

In Fig.~\ref{fPa} we plot some numerical results both for $\hat p^{(\nu)}_{N,n,\alpha}$ and for the $q$--logarithm
\begin{equation}\label{w}
w^{(\nu)}_{\alpha}(x)\coloneqq\log_{q_\alpha(\nu)}\frac{P_{q_\alpha(\nu)}(x)}{P_{q_\alpha(\nu)}(0)}=-x^2,
\end{equation}
comparing with the theoretical predictions. Observe that the following relation between $q_\alpha(\nu)$ and $q_1(\nu)$ holds:
\begin{equation}\label{Aalgebra}
\frac{1}{1-q_\alpha(\nu)}=\frac{1}{1-q_1(\nu)}+1.
\end{equation}

\paragraph{Entropy}
Let us now focus on which entropic functional is extensive for the above model. A natural candidate is in principle the nonadditive entropy  \cite{tsallis1988} 
\beq
S_{q_\text{ent}}^{(\nu,\alpha)}\coloneqq \frac{1-\sum_{n=0}^{N}\binom{N}{n}\left(\hat r^{(\nu)}_{N,n,\alpha}\right)^{q_\text{ent}}}{1-q_\text{ent}}.\label{Tsallis}
\eeq
Using a generalized entropic form \cite{hanel2011a,tempesta2011,tempesta2014}, including the nonadditive entropy as particular case, in \cite{hanel2011} it has been shown that for a wide class of triangles, the number of microstates $\Omega$, as a function of the system size $N$, increases according to the law $\Omega(N)=2^{N}$. This leads to a scenario in which the only possible value of $q_{\text{ent}}$ making the entropy \eqref{Tsallis} extensive for $\alpha=1$ is $q_{\text{ent}}=1$, which corresponds to the Boltzmann--Gibbs entropy. 
%Instead, the extensivity of the nonadditive entropy for $q \neq 1$ is guaranteed in those cases when the occupancy of phase space is not complete, i.e.  fully dimensional subsets are forbidden. 
Similar arguments, based on the Laplace--de Finetti theorem, also yield $q_{\text{ent}}=1$ \cite{hanel2009}.

\begin{figure}
\centering
\includegraphics{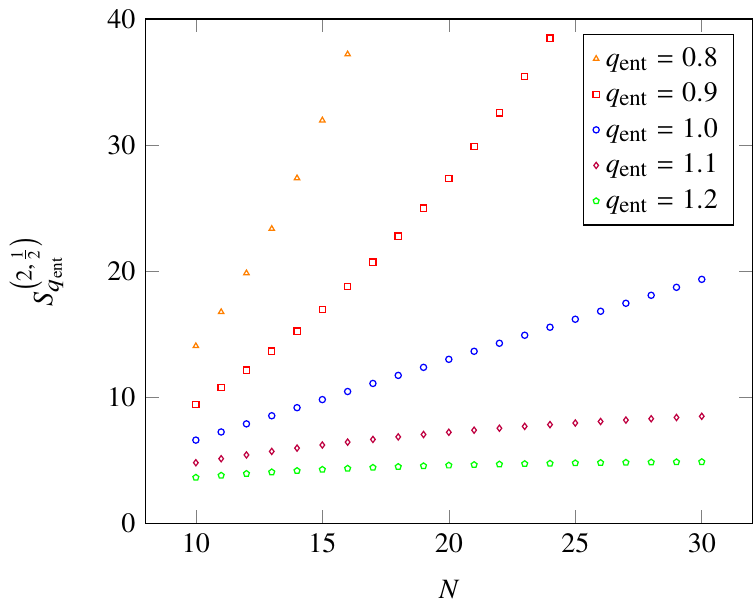}
\caption{Values of the $S^{\left(2,\alpha\right)}_{q_\text{ent}}$ entropy \eqref{Tsallis} for $\alpha=\frac{1}{2}$ versus $N$ for different values of $q_\text{ent}$. \label{S_q_nu2_alpha}}
\end{figure}

The same kind of reasoning of \cite{hanel2011,hanel2009} applies for the class of models we address here, and, therefore, also here we expect  $q_{\text{ent}}=1$. This has been confirmed by a numerical analysis. In Fig.~\ref{S_q_nu2_alpha}, the $q$--entropy \eqref{Tsallis} is plotted as a function of $N$ for the particular case $\nu=2$ and $\alpha=\frac{1}{2}$. In agreement with the above, we find that the value of $q_\text{ent}$ which makes the entropy $S_{q_\text{ent}}^{(\nu,\alpha)}$ extensive is indeed $q_{\text{ent}}=1$.

\subsection{The $\beta$--numbers and the $\beta$--triangles}
Let us consider another deformation of the generalized Leibniz triangles introduced in Section \ref{LT}.
\begin{dbeta}
Given $n\in\mathds{N}\cup{0}$, and $\beta>0$, $\beta\neq 1$, we shall call {\it$\beta$--numbers} the real numbers defined as follows:
\bea
[n]_\beta\coloneqq\begin{cases}0&\text{if $n=0$},\\n\left(1-\frac{1-\beta}{1-{\beta}^n}\right)+1\equiv  n\, b_{n}(\beta)& \text{if $n\geq 1$},\end{cases}
\label{beta_number}
\eea
where we introduced
\begin{equation}
b_n(\beta)\coloneqq 1+\frac{1}{n}-\frac{1-\beta}{1-{\beta}^n}.
\end{equation}
\end{dbeta}
Note that, so defined, the $\beta$--numbers are related to the $\alpha$--numbers as
\begin{equation}
[n]_\beta=\{n-1\}_\beta+1.
\end{equation}
It follows that $[n]_1\coloneqq\lim_{\beta\to 1}[n]_\beta=n$, hence the $\beta\to1$ limit recovers the ordinary numbers. The \textit{$\beta$--factorial} can now be defined as
\begin{equation}
[0]_\beta!\coloneqq 1,\quad  [n]_\beta!\coloneqq\prod_{k=1}^{n}[k]_\beta=n!\prod_{k=1}^nb_k(\beta),\quad n\geq 1.
\end{equation}
The factorial number is recovered as $[n]_1!\coloneqq \lim_{\beta\to 1}[n]_\beta!=n!$. We can further define the $\beta$--binomial coefficient as
\begin{multline}
\begin{bmatrix}N\\n\end{bmatrix}_\beta\coloneqq\frac{[N]_\beta!}{[n]_\beta![N-n]_\beta!}\\=\begin{cases}1&\text{if $n=0$ or $n=N$},\\\binom{N}{n}\prod_{k=1}^{N-n}\frac{b_{k+n}(\beta)}{b_{k}(\beta)}&\text{if $1\leq n\leq N-1$}.\end{cases}
\end{multline}
where, as expected, $\lim_{\beta\to 1}\left[\begin{smallmatrix}N\\n\end{smallmatrix}\right]_\beta=\binom{N}{n}$.

We can introduce therefore the {\em Leibniz--like $\beta$--triangle} as
\begin{equation}
r^{(1)}_{N,n,\beta}\coloneqq\dfrac{1}{[N+1]_\beta} \dfrac{1}{\left[\begin{smallmatrix}N\\n\end{smallmatrix}\right]_\beta},\quad  n=0,\dots,N. \label{beta-Leibniz}
\end{equation}
The Leibniz triangle \eqref{Leibniz_triangle} is obtained in the $\beta\to 1$ limit, $\lim_{\beta\to 1}r_{N,n,\beta}^{(1)}\equiv r_{N,n}^{(1)}$. As before, in order to get a set of probabilities we have to normalize the triangles, obtaining
\begin{equation}
\hat r^{(1)}_{N,n,\beta}\coloneqq\dfrac{r^{(1)}_{N,n,\beta}}{\sum_{k=0}^N\binom{N}{k}r^{(1)}_{N,k,\beta}},
\label{triangulo_normalizado_beta}
\end{equation}
whose associated probabilities
\begin{equation}
\hat p^{(1)}_{N,n,\beta}\coloneqq\binom{N}{n}\hat r^{(1)}_{N,n,\beta}
\label{probabilidad_normalizada_beta}\end{equation}
satisfy by construction the normalization condition $\sum_{n=0}^{N}\hat{p}^{(1)}_{N,n,\beta}=1$. 

We will now generalize the Leibniz $\beta$--triangle by properly substracting subtriangles of it. In analogy with the previous deformation, let us now introduce a two--parameter family of triangles
\begin{equation}
r^{(\nu)}_{N,n,\beta}\coloneqq\frac{r^{(1)}_{N+2(\nu-1),n+\nu-1,\beta}}{r^{(1)}_{2(\nu-1),\nu-1,\beta}}\equiv\mu^{(\nu)}_{N,n,\beta}r^{(\nu)}_{N,n},\label{r_nu_beta}
\end{equation}
where $\nu\in\mathds N$ and
\begin{equation}\label{MuBeta}
\mu^{(\nu)}_{N,n,\beta}\coloneqq \frac{b_{2\nu-1}(\beta)}{b_{N+2\nu-1}(\beta)}\prod_{k=1}^{\nu-1}\frac{b_{k+\nu-1}(\beta)}{b_k(\beta)}\prod_{k=1}^{N-n+\nu-1}\frac{b_k(\beta)}{b_{k+n+\nu-1}(\beta)}.
\end{equation}
The $\beta\to 1$ limit of the two--parameters family of triangles \eqref{r_nu_beta} yields the undeformed family \eqref{r_nu}, $\lim_{\beta\to 1}r^{(\nu)}_{N,n,\beta}\equiv r^{(\nu)}_{N,n}$. After the needed normalization, we obtain the family $\hat r_{N,n,\beta}^{(\nu)}$ and the corresponding probabilities
\begin{subequations}\label{beta_normalizado}\begin{align}
\hat r_{N,n,\beta}^{(\nu)}&\coloneqq\frac{r_{N,n,\beta}^{(\nu)}}{\sum_{k=0}^N\binom{N}{k}r_{N,k,\beta}^{(\nu)}},\label{r_nu_beta_normalizado}\\
\hat p_{N,n,\beta}^{(\nu)}&\coloneqq\binom{N}{n}\hat r_{N,n,\beta}^{(\nu)}.\label{P_nu_beta_normalizado}
\end{align}
\end{subequations}
It can be proved that the triangle \eqref{r_nu_beta_normalizado}  is asymptotically scale--invariant,
\begin{equation}\lim_{\substack{N\to\infty\\\frac{n}{N}\equiv\eta\ \text{fixed}}}\frac{\hat r^{(\nu)}_{N-1,n,\beta}}{\hat r^{(\nu)}_{N,n, \beta}+\hat r^{(\nu)}_{N,n+1, \beta}}=1.\label{limite_beta}\end{equation}

The $\beta$--numbers appear as a variation of $\alpha$--numbers and, moreover, they have the same asymptotic behavior, $\lim_{n\to\infty}\frac{[n]_\alpha}{\{n\}_\alpha}=1$. Therefore, it could be expected that the behavior of $\hat p^{(\nu)}_{N,n,\alpha}$ is the same of $\hat p^{(\nu)}_{N,n,\beta}$ for $N\to\infty$ with respect to the parameters of the deformation. However this is true only when we consider values of the parameters greater than one. Indeed, for $\beta$--triangles, the following theorem holds:
\begin{thbeta}\label{ThBeta}
The family of triangles \eqref{beta_normalizado} with $\nu\in\mathds N$ and $\beta>0$ satisfies the property
\begin{equation}
\frac{N}{2\sqrt{\nu-\chi(\beta)}}\hat p_{N,n,\beta}^{(\nu)} \asym{N\gg 1} P_{q_\beta(\nu)}(x),\label{EqThB}
\end{equation}
where we have introduced the function
\begin{equation}\label{chi}
\chi(\beta)\coloneqq 1+\delta_{\beta,1}-\max\left\{1,\frac{1}{\beta}\right\}=\begin{cases}0&\text{for $\beta>1$},\\1&\text{for $\beta=1$},\\1-\frac{1}{\beta}&\text{for $0<\beta<1$,} \end{cases}
\end{equation}
and the properly centered and rescaled variable
\begin{equation}\label{cov}
x\coloneqq 2\sqrt{\nu-\chi(\beta)}\left(\frac{n}{N}-\frac{1}{2}\right).
\end{equation}
In Eq.~\eqref{EqThB},  $P_{q_\beta(\nu)}(x)$ is a $q$--Gaussian with 
\begin{equation}\label{qbeta}
q_\beta(\nu)=1-\frac{1}{\nu-\chi(\beta)}=\begin{cases}1-\frac{1}{\nu}&\text{for $\beta>1$},\\1-\frac{1}{\nu-1}&\text{for $\beta=1$},\\1-\frac{\beta}{\beta\nu+1-\beta}&\text{for $0<\beta<1$.} \end{cases}\end{equation}
\end{thbeta}
%\begin{proof} 
See Appendix \ref{A-beta} for the proof. \\
%\end{proof}

\begin{figure*}
{\centering
\includegraphics{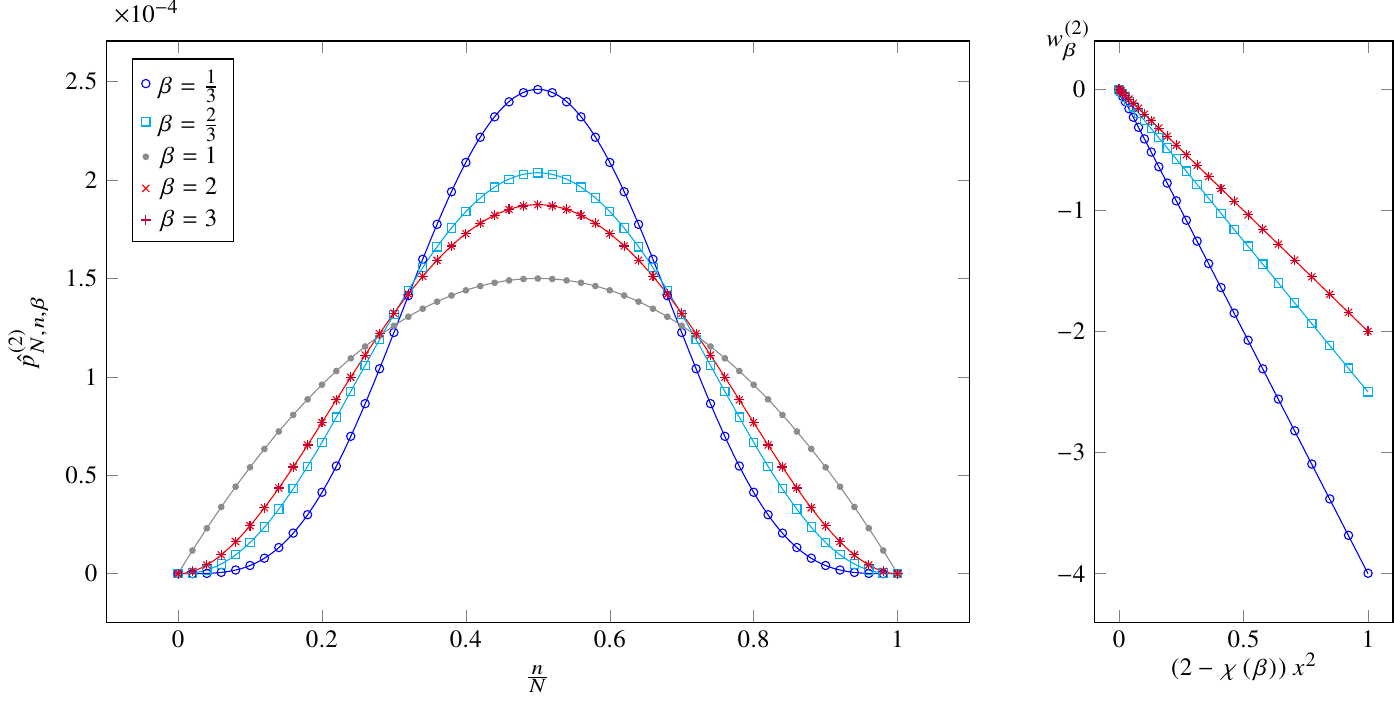}}
\caption{Probability distributions \eqref{P_nu_beta_normalizado} and $w_\beta^{(\nu)}(x)\coloneqq\log_{q_\beta(\nu)}\frac{P_{q_\beta(\nu)}(x)}{P_{q_\beta(\nu)}(0)}$ for $\nu=2$, $N=10^4$ and different values of $\beta$; the derived asymptotic behavior is also depicted as continuous line for each value of $\beta$: for the sake of clarity not all data are plotted. The corresponding values of $q_\beta(2)$ are given by Eq.~\eqref{qbeta}.\label{fPb}}
\end{figure*}

In Fig.~\ref{fPb} we present some numerical results for $\beta$--triangles for different values of $\beta$. We plot also
\begin{equation}
w_\beta^{(\nu)}(x)\coloneqq\log_{q_\beta(\nu)}\frac{P_{q_\beta(\nu)}(x)}{P_{q_\beta(\nu)}(0)}
\end{equation}
for $\nu=2$ and different values of $\beta$. 

Observe that $q_\alpha(\nu)=q_\beta(\nu)$ when $\alpha>1$ and $\beta>1$; moreover we can write the following relation between the limiting value $q_\beta(\nu)$ for the deformed triangles ($\beta\neq 1$) and the limiting value $q_1(\nu)$ for the undeformed triangle:
\begin{equation}\label{algebra}
\frac{1}{1-q_\beta(\nu)}=\frac{1}{1-q_1(\nu)}+1-\chi(\beta).
\end{equation}
In particular
\begin{equation}
\lim_{\beta\to 1^\pm}\frac{1}{1-q_\beta(\nu)}=\frac{1}{1-q_1(\nu)}+1.
\end{equation}
Eq.~\eqref{algebra} can be written, for $\beta\neq 1$, as
\begin{equation}
\frac{\min\{\beta,1\}}{1-q_\beta(\nu)}=\frac{\min\{\beta,1\}}{1-q_1(\nu)}+1.\label{Balgebra}
\end{equation}
%The different functional dependence of $q_\beta$ on the parameter $\beta$ in the two intervals $(0,1)$ and $(1,+\infty)$ can be found in other probabilistic models in the literature (for example, \citet{marsh2006} analyzed a probabilistic model in which $q_\text{ent}$ has a similar behavior with respect to a certain parameter of the model). 

Finally, following the same arguments adopted for the $\alpha$--triangles, it can be easily verified that also in this case we have $q_{\text{ent}}=1$.

\section{A non asymptotically scale--invariant deformation}

Inspired by the \textit{$Q$--calculus} \cite{kac2002}, we shall consider an alternative deformation of the Leibniz triangle based on the so called {\em $Q$--numbers}, defined as:
\begin{equation}
\llbracket n\rrbracket_{\gamma}\coloneqq\frac{1-\gamma^n}{1-\gamma},\quad\gamma\in(0,\infty)\setminus\{1\},\label{q-number}
\end{equation}
where we have used the notations $\gamma$ and $\llbracket n\rrbracket_{\gamma}$ instead of the usual ones in order to avoid confusion with the entropic index $q$ in nonextensive statistical mechanics and the previously introduced $\alpha$--numbers and $\beta$--numbers. For this reason, in the remainder of the paper we will refer to the $Q$--numbers as the \textit{$\gamma$--numbers}.

Note that $\gamma$--numbers \eqref{q-number}, $\alpha$--numbers \eqref{alpha_number} and $\beta$--numbers \eqref{beta_number} are related as follows:
\begin{equation}
[n]_\gamma=\{n-1\}_\gamma+1=n\left(1-\frac{1}{\llbracket n\rrbracket_{\gamma}}\right)+1.
\end{equation} 
In the limit $\gamma\to 1$ the $\gamma$--numbers reduce to the ordinary numbers, $\lim_{\gamma\to 1}[\![n]\!]_{\gamma}=n$. Moreover, a {\em$\gamma$--factorial} can be defined as
\begin{equation}\llbracket 0\rrbracket_{\gamma}!\coloneqq 1,\qquad
\llbracket n\rrbracket_{\gamma}!\coloneqq\prod_{k=1}^{n}\llbracket k \rrbracket_{\gamma}, \quad \lim_{\gamma\to 1}\llbracket n\rrbracket_\gamma!=n!,\quad n\geq 1,
\end{equation}
as well as the standard {\em Gauss binomial coefficients}
\begin{multline}
\left\llbracket\begin{matrix}N\\n
\end{matrix}\right\rrbracket_{\gamma}\coloneqq\frac{\llbracket N\rrbracket_{\gamma}!}{\llbracket n\rrbracket_{\gamma}!\llbracket N-n\rrbracket_{\gamma}!}\\=\begin{cases}1&\text{if $n=0$ or $n=N$,}\\\prod_{k=1}^{N-n}\frac{1-\gamma^{k+n}}{1-\gamma^k}&\text{if $0<n<N$},\end{cases}
\end{multline}
with $\lim_{\gamma\to 1}\left\llbracket\begin{smallmatrix}N\\n
\end{smallmatrix}\right\rrbracket_{\gamma}=\binom{N}{n}$. The Gauss binomial coefficients can be arranged in a {\em Pascal $\gamma$--triangle}:
\begin{center}
\includegraphics{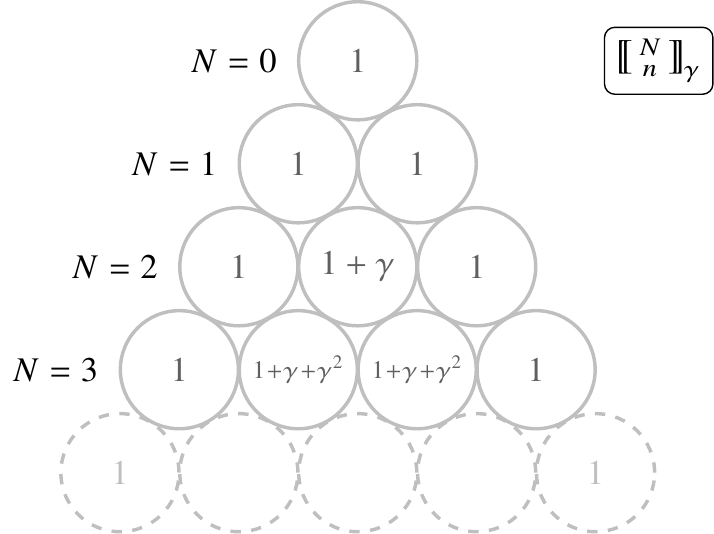}
\end{center}
which reduces to the Pascal triangle for $\gamma\to 1$. We define the {\em Leibniz--like $\gamma$--triangle} as:
\begin{equation}
r^{(1)}_{N,n,\gamma}\coloneqq\dfrac{1}{\llbracket N+1\rrbracket_{\gamma}}
\dfrac{1}{{\left\llbracket\begin{smallmatrix}N\\n
\end{smallmatrix}\right\rrbracket}_{\gamma}},
\label{Leibniz_q_triangle}
\end{equation}
with corresponding normalized triangle
\begin{equation}
\hat r^{(1)}_{N,n,\gamma}\coloneqq\dfrac{r^{(1)}_{N,n,\gamma}}{\sum_{k=0}^N\binom{N}{k}r^{(1)}_{N,k,\gamma}},\label{Leibniz_q_triangle_normalizado}
\end{equation}
and associated probabilities
\begin{equation}
\hat p^{(1)}_{N,n,\gamma}\coloneqq\binom{N}{n}\hat r^{(1)}_{N,n,\gamma}.
\label{probabilidad_q_Leibniz}\end{equation}
In the limit $\gamma\to 1$, triangles \eqref{Leibniz_q_triangle} and \eqref{Leibniz_q_triangle_normalizado} reduce to the Leibniz triangle. 

Proceeding as in the previous Section, we can introduce the family
\begin{equation}
r^{(\nu)}_{N,n,\gamma}\coloneqq\frac{r^{(1)}_{N+2(\nu-1),n+\nu-1,\gamma}}{r^{(1)}_{2(\nu-1),\nu-1,\gamma}},
\end{equation}
deformation of the family \eqref{r_nu}, and the corresponding normalized version and its associated probability
\begin{subequations}
\begin{align}
\hat r^{(\nu)}_{N,n,\gamma}&\coloneqq \frac{r^{(\nu)}_{N,n,\gamma}}{\sum_{k=0}^N\binom{N}{k}r^{(\nu)}_{N,k,\gamma}},\label{LeibnizNuqNormalizado}\\
\hat p^{(\nu)}_{N,n,\gamma}&\coloneqq \binom{N}{n} \hat r^{(\nu)}_{N,n,\gamma}.
\end{align}
\end{subequations}
Remarkably, the \textit{$\gamma$--triangles} \eqref{LeibnizNuqNormalizado} are {\em neither strictly nor asymptotically} scale--invariant since
\begin{equation}\lim_{\substack{N\to\infty\\\frac{n}{N}\equiv \eta\text{ fixed}}}\frac{\hat r^{(\nu)}_{N-1,n,\gamma}}{\hat r^{(\nu)}_{N,n,\gamma}+\hat r^{(\nu)}_{N,n+1, \gamma}}=\begin{cases}
\frac{1}{2}&\text{for $\gamma<1$,}\\0&\text{for $\gamma>1$.}\end{cases}\end{equation}

In addition, probabilities \eqref{probabilidad_q_Leibniz} do {\em not} approach $q$--Gaussians with $q\neq 1$, as limiting distributions for large values of $N$.

\begin{figure*}
\subfloat[Probability distribution $\hat p^{(\nu)}_{N,n,\gamma}$ for $\nu=2$ and different values of $\gamma$ and theoretical prediction for $N=10^4$: not all data are represented for sake of clarity.\label{qm1}]{\centering
\includegraphics{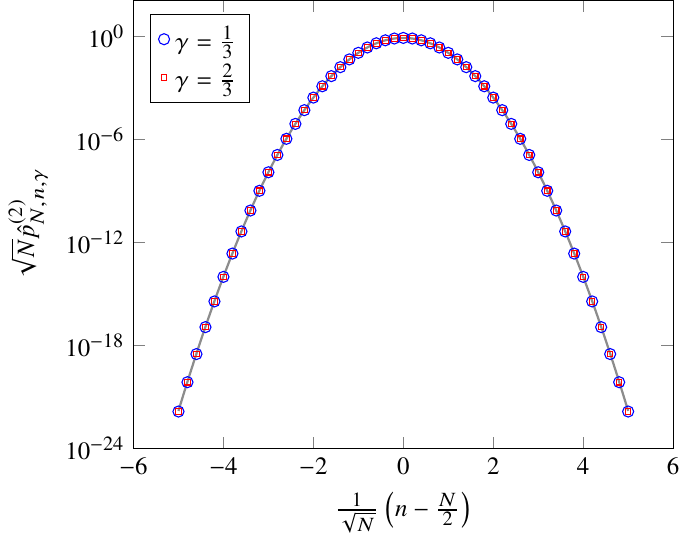}}\hspace{0.5cm}
\subfloat[Probability distribution $\hat p^{(2)}_{N,n,\frac{3}{2}}$ for different values of $N$.\label{qM1}]{\centering \includegraphics{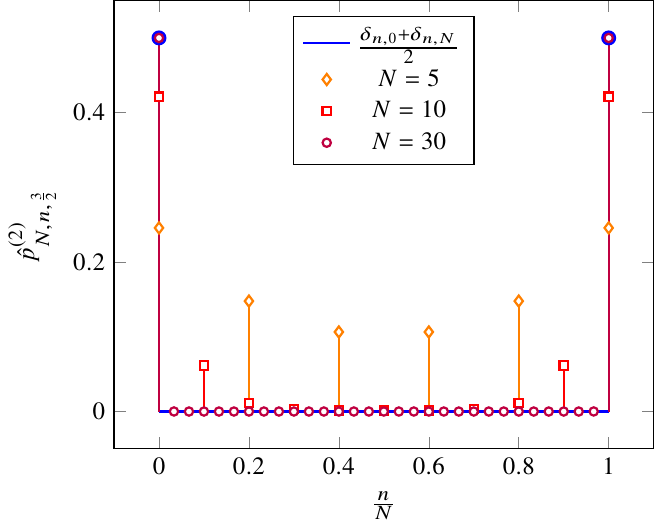}}
\caption{Numerical results for $\gamma$--triangles.}
\end{figure*}

Indeed, let us start from $\gamma\in(0,1)$: for $N\gg 1$, $\nu>1$ and $\frac{n}{N}\equiv\eta$ fixed. Then
\begin{multline}
r^{(\nu)}_{N,n,\gamma}=\frac{1}{r^{(1)}_{2(\nu-1),\nu-1,\gamma}}\frac{1-\gamma}{1-\gamma^{N+2\nu-1}}\cdot\\\cdot\prod_{k=1}^{N(1-\eta)+\nu-1}\frac{1-\gamma^k}{1-\gamma^{k+N\eta+\nu-1}}%\sim
%\frac{(1-\gamma)\prod_{k=1}^{N(1-\eta)}\left(1-\gamma^k\right)}{r^{(1)}_{2(\nu-1),\nu-1,\gamma}}
%\\=\frac{(1-\gamma)\prod_{k=1}^{\infty}\left(1-\gamma^k \right )}{r^{(1)}_{2(\nu-1),\nu-1,\gamma}}\left(1+\gamma^{N(1-\eta)}\right)
\\\sim\frac{(1-\gamma)\prod_{k=1}^{\infty}\left(1-\gamma^k \right )}{r^{(1)}_{2(\nu-1),\nu-1,\gamma}}\sim O(1).
\end{multline}
The relevant contribution in the probability distribution shape is therefore simply given by the binomial coefficient, and therefore, for $N\gg 1$, we recover the Gaussian distribution (see Fig.~\ref{qm1}) for all values of $\gamma\in(0,1)$:
\begin{equation}
\hat p^{(\nu)}_{N,n,\gamma}\asym{N\gg 1}\sqrt{\frac{2}{\pi N}}e^{-2N\left(\eta-\frac{1}{2}\right)^2}.
\end{equation}
For $\gamma>1$ we can evaluate the asymptotic distribution as well. We can write
\begin{equation}
\frac{1}{\hat p^{(\nu)}_{N,n,\gamma}}=\frac{1}{\binom{N}{n}}\sum_{k=0}^Nh_k(n,N;\gamma) \,,
\end{equation}
where
\begin{equation}\textstyle
h_k(n,N;\gamma)\coloneqq\begin{cases}\binom{N}{k}\prod^{k-n}_{j=1}\frac{1-\gamma^{n+\nu-1+j}}{1-\gamma^{N-k+\nu-1+j}}&\text{for $k>n$,}\\\binom{N}{n}&\text{for $k=n$,}\\\binom{N}{k}\prod^{n-k}_{j=1}\frac{1-\gamma^{N-n+\nu-1+j}}{1-\gamma^{k+\nu-1+j}}&\text{for $k<n$.}\end{cases}
\end{equation}
For $N\gg 1$, $\frac{n}{N}\equiv \eta$, and denoting by $\frac{k}{N}\equiv \kappa$, we have that
\begin{equation}\textstyle
h_{N\kappa}(N\eta,N;\gamma)\sim\binom{N}{N\kappa}\gamma^{N^2\left(\eta-\kappa\right)\left(1-\eta-\kappa\right)}.
\end{equation}
It is easily seen that the quantity $\sum_{N\kappa}\frac{h_{N\kappa}}{\binom{N}{n}}$ can be finite only for $\eta=0$ or $\eta=1$, otherwise at least one of its addends diverges and $\hat p_{N,n,\gamma}^{(\nu)}\xrightarrow{N\to\infty}0$, $n=1,\dots,N-1$. Finally we get (see Fig.~\ref{qM1} for a numerical comparison)
\begin{equation}
\hat p_{N,n,\gamma}^{(\nu)}\asym{N\gg 1}\frac{\delta_{n,0}+\delta_{n,N}}{2}.
\end{equation}

\section{Final remarks}
\begin{figure*}
\subfloat[Function $q_\alpha(2)$ and its discontinuity for $\alpha=1$, see Eq.~\eqref{qalpha}.]{
\includegraphics{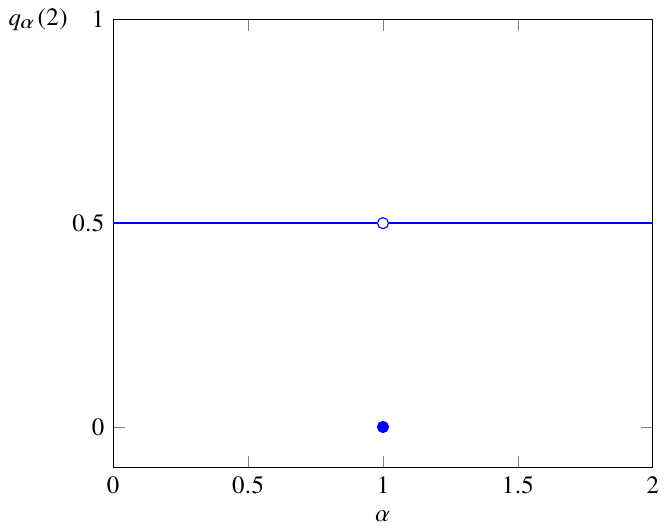}}\hspace{0.3cm}
\subfloat[Function $q_\beta(2)$ and its discontinuity for $\beta=1$, see Eq.~\eqref{qbeta}.]{
\includegraphics{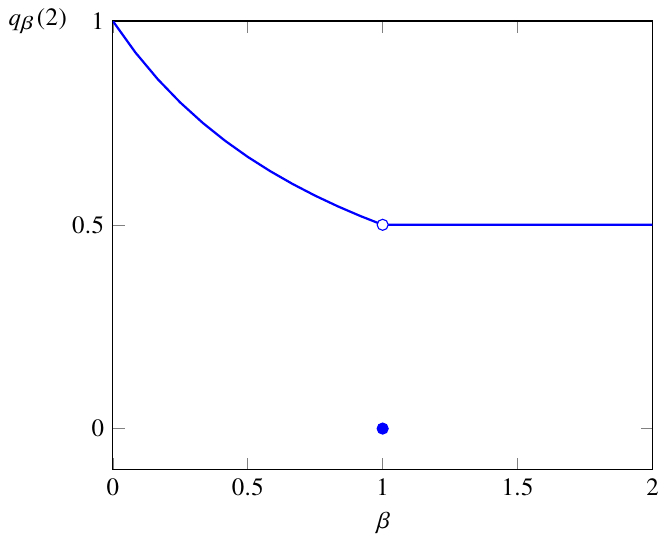}}
\caption{Functions $q_\alpha(\nu)$ and $q_\beta(\nu)$ evaluated for $\nu=2$ and different values of $\alpha$ and $\beta$ respectively.\label{fqAB}}
\end{figure*}
\begin{figure}\centering
\includegraphics{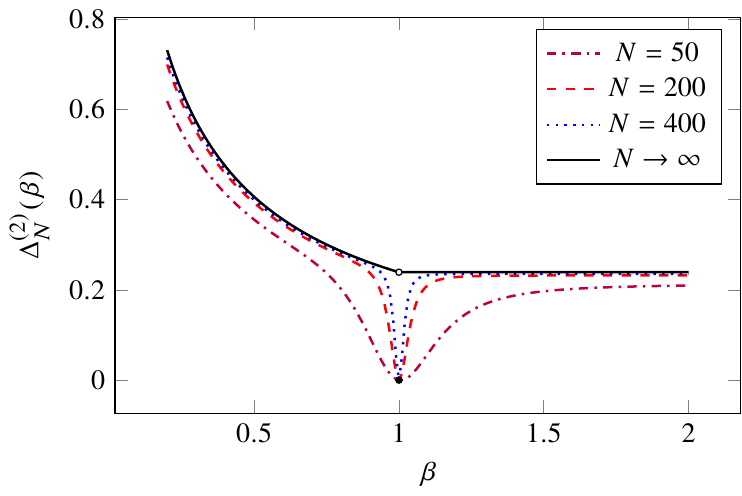}
\caption{Values of $\Delta_N^{(\nu)}(\beta)$ for $\nu=2$ and different values of $N$: the limiting function $\lim_{N\to\infty}\Delta_N^{(2)}(\beta)$ is also represented, including its discontinuity for $\beta=1$.\label{fdistanza}}
\end{figure}

We introduced three deformations --- the $\alpha$--, the $\beta$-- and the $\gamma$--triangles --- of the family of triangles proposed in \cite{rodriguez2008}, in order to analyze the robustness of the $q$--Gaussian family as attractors. Each one of the three proposed deformations depends on a single parameter in such a way that the undeformed family is recovered when the value of that parameter equals $1$. We observed that, in all considered cases, the limiting distribution changes abruptly with respect to the undeformed case when a deformed triangle is considered. Moreover, asymptotically equivalent deformations of natural numbers (the $\alpha$--deformation and the $\beta$--deformation) exhibit different limiting behaviors, thus illustrating the high sensitivity of the limiting distribution with respect to the exact form of the deformation. However, for the considered \textit{asymptotically scale--invariant deformations}, the limiting distribution of the deformed triangles \textit{is still a $q$--Gaussian}, with a different value of $q$ generically depending on the parameter of the deformation. Moreover, a discontinuity appears in $q$ as function of the parameters $\alpha,\beta$ of the deformations, in correspondence of the undeformed case $\alpha=\beta=1$ (see Fig.~\ref{fqAB}). In particular
\begin{equation}
\lim_{\alpha\to 1}\frac{1}{1-q_\alpha(\nu)}=\lim_{\beta\to 1}\frac{1}{1-q_\beta(\nu)}=\frac{1}{1-q_1(\nu)}+1.
\end{equation}
Remarkably, this discontinuity appears when we switch from a scale--invariant triangle to an \textit{asymptotically} scale--invariant triangle. For $\beta$--triangles, and similarly for $\alpha$--triangles, this discontinuity expresses the fact that
\begin{equation}
\lim_{\beta\to 1}\lim_{N\to\infty}N \hat p^{(\nu)}_{N,n,\beta}\neq \lim_{N\to\infty}\lim_{\beta\to 1}N \hat p^{(\nu)}_{N,n,\beta}.
\end{equation}
To exemplify this, let us introduce the following function:
\begin{equation}
\Delta_N^{(\nu)}(\beta)\coloneqq \sqrt{N\sum_{n=0}^N\left|\hat p_{N,n,\beta}^{(\nu)}-p_{N,n}^{(\nu)}\right|^2}.
\end{equation}
The quantity $\Delta_N^{(\nu)}(\beta)$ is defined in such a way that, for $N\to\infty$, it remains finite. Indeed, $N\hat p_{N,n,\beta}^{(\nu)}=O(1)$ in the $N\to\infty$ limit, and therefore $\sum_{n=0}^N|\hat p_{N,n,\beta}^{(\nu)}-p_{N,n}^{(\nu)}|^2=O\left(\frac{1}{N}\right)$. In Fig.~\eqref{fdistanza} it can be seen that the convergence of $\Delta_N^{(\nu)}(\beta)$ to $\Delta_\infty^{(\nu)}(\beta)\coloneqq\lim_{N\to\infty}\Delta_N^{(\nu)}(\beta)$ is not uniform and that a discontinuity for $\beta=1$ appears in the $N\to\infty$ limit.

The $\alpha$--triangle is strictly stable under the action of the $\alpha$--deformation for $\alpha\neq 1$, whereas a dependence of the limiting value of $q$ on $\beta$ (for $\beta<1$) appears when the $\beta$--deformation is considered, although the limiting distribution is still a $q$--Gaussian (\citet{marsh2006} analyzed a probabilistic model in which $q_\text{ent}$ has a similar behavior with respect to a certain parameter of the model). The structure of relations \eqref{Aalgebra} and \eqref{Balgebra} is quite common in the literature: it has been observed, indeed, that for many systems characterized by a set of values of $q$, $\{q_m\}_{m\in M}$, $M\coloneqq\{0,1,\dots\}\subseteq \mathds Z$, a permutation of the indices $m$ can be found such that $\forall m\in M$ and $\bar\alpha\in(0,2]$ \cite{tsallis2009bjp}%we can express every value $q_m$ in term of a special element $q_0$ of the set as
\begin{equation}
\frac{\bar\alpha}{1-q_m}=\frac{\bar\alpha}{1-q_0}+m.
\end{equation}

Finally, using the $\gamma$--deformation, that is {\it not} an asymptotically scale--invariant deformation,  for $\gamma>1$ we obtain a limiting distribution that is \textit{not} a $q$--Gaussian distribution, and for $\gamma <1$ we always obtain a Gaussian. This fact suggests that the (asymptotic) scale--invariance property plays a central role in the robustness of the set of $q$--Gaussian distributions as limiting distributions. More specifically, the set of $q$--Gaussians appears to be \textit{robust} under asymptotically scale--invariant deformations. It may be interesting to investigate further the role of asymptotically scale--invariant deformations in the stability of the $q$--Gaussian limiting distributions, to properly identify the conditions under which the basin of $q$--Gaussians is an attractor for these probabilistic models. This line of research could ultimately provide a deeper understanding of why there are so many $q$--Gaussians and $q$--exponentials in Nature.

\section*{Acknowledgments}
We acknowledge partial support from CNPq and FAPERJ (Brazilian agencies). G.\,S. and C.\,T. also acknowledge the financial support of the John Templeton Foundation. The research of P.\,T. has been supported by the grant FIS2011--22566, Ministerio de Ciencia e Innovaci\'on, Spain. A.\,R. thanks financial support from DGU-MEC (Spanish Ministry of Education) through project PHB2007-0095-PC and Comunidad de Madrid through project MODELICO.  

\appendix
\section{Asymptotic behavior of $\alpha$--triangles}\label{A-alpha}
In this Appendix we prove Theorem \ref{ThAlpha} on the asymptotic distribution of $\alpha$--triangles. We restate the theorem here for the reader's convenience.
\begin{thalpha2}
The family of triangles \eqref{r_nu_alpha} with $\nu\in(1,+\infty)$ and $\alpha\in\mathds R^+\setminus\{0\}$ satisfies the property
\begin{equation}\label{EqThA2}
\frac{N}{2\sqrt{\nu-\delta_{\alpha,1}}}\hat p_{N,n,\alpha}^{(\nu)} \asym{N\gg 1} P_{q_\alpha(\nu)}(x),
\end{equation}
where we have introduced the properly centered and rescaled variable
\begin{equation}\label{xvar2}
x\coloneqq 2\sqrt{\nu-\delta_{\alpha,1}}\left(\frac{n}{N}-\frac{1}{2}\right).
\end{equation}
In Eq.~\eqref{EqThA2}, $P_{q_\alpha(\nu)}(x)$ is a $q$--Gaussian with 
\begin{equation}
q_\alpha(\nu)\coloneqq 1-\frac{1}{\nu-\delta_{\alpha,1}}=\begin{cases}1-\frac{1}{\nu}&\text{for $\alpha\neq 1$},\\1-\frac{1}{\nu-1}&\text{for $\alpha=1$.} \end{cases}\end{equation}
\end{thalpha2}

\begin{proof} The strategy of the first part of the proof consists in a generalization of the argument in \cite{rodriguez2008}. We want to evaluate the asymptotic behavior, for $N\gg 1$, of
\begin{equation}
\binom{N}{n}r^{(\nu)}_{N,n,\alpha}
\end{equation}
with $\eta\coloneqq\frac{n}{N}$ fixed.  Let us firstly observe that
\begin{equation}
r^{(\nu)}_{N,0}=\frac{\Gamma(2\nu)\Gamma(N+\nu)}{\Gamma(\nu)\Gamma(N+2\nu)}=\frac{\Gamma(2\nu)}{\Gamma^2(\nu)}\sum_{k=0}^\infty\frac{(1-\nu)_{k}}{k!(N+\nu+k)},
\end{equation}
where we used the formula for the beta function\footnote{This formula can be obtained using the Newton's generalized binomial theorem on the expression $\mathrm B(a,b)\coloneqq \int_0^1t^{a-1}(1-t)^{b-1}dt$.}
\begin{equation}
\mathrm B(a,b)\coloneqq\frac{\Gamma(a)\Gamma(b)}{\Gamma(a+b)}=\int_0^1t^{a-1}(1-t)^{b-1}dt=\sum_{k=0}^\infty\frac{(1-b)_{k}}{k!(a+k)},
\end{equation}
and $(a)_k\coloneqq a(a+1)\dots(a+k-1)$ is the rising factorial.
The previous formula allows us to write down a general expression for the element $r^{(\nu)}_{N,n}$ as follows:
\begin{multline}
r^{(\nu)}_{N,n}=\sum_{i=0}^n(-1)^{n-i}\binom{n}{i}r^{(\nu)}_{N-i,0}\\
=\frac{\Gamma(2\nu)}{\Gamma^2(\nu)}\sum_{i=0}^n\sum_{k=0}^\infty \frac{(-1)^{n-i}(1-\nu)_{k}}{k!}\binom{n}{i}\frac{1}{N+\nu-i+k}.
\end{multline}
To evaluate the large $N$ behavior, we use the saddle point approximation,
\begin{multline}
\sum_{i=0}^n (-1)^{n-i}\binom{n}{i}\frac{1}{N+\nu-i+k}\\=\int_0^{\infty}\left(e^\xi-1\right)^n e^{-(N+\nu+k)\xi}d\xi\\\sim \sqrt{\frac{2\pi}{N}}\left(1-\eta\right)^{\nu+k+N(1-\eta)-\frac{1}{2}}\eta^{N\eta+\frac{1}{2}},
\end{multline}
where $\eta\coloneqq\frac{n}{N}$. Inserting the previous term in the complete expression we have
\begin{equation}
r_{N,n}^{(\nu)}\sim\frac{\Gamma(2\nu)}{\Gamma^2(\nu)}\sqrt{\frac{2\pi}{N}}\left(1-\eta\right)^{\nu+N(1-\eta)-\frac{1}{2}}\eta^{\nu+N\eta-\frac{1}{2}}
\end{equation}
Moreover, using the Stirling formula,
\begin{multline}
N\binom{N}{n}\asym{N\gg 1}\sqrt{\frac{N}{2\pi}}(1-\eta)^{-N(1-\eta)-\frac{1}{2}}\eta^{-N\eta-\frac{1}{2}}\\\Rightarrow
N\binom{N}{n}r^{(\nu)}_{N,n}\asym{N\gg 1}\frac{\Gamma(2\nu)}{\Gamma^2(\nu)}\left[\left(1-\eta\right)\eta\right]^{\nu-1}.
\end{multline}
To complete the proof, we only need to evaluate the asymptotic behavior of $\mu^{(\nu)}_{N,n,\alpha}$ for large $N$ at fixed $\eta$. Obviously, $\lim_{\alpha\to 1}\mu^{(\nu)}_{N,n,\alpha}=1$. For $\alpha\neq 1$ we have that, by direct computation on the expression \eqref{MuAlpha}
\begin{equation}
\mu^{(\nu)}_{N,n,\alpha}\asym{N\gg 1}N\eta(1-\eta)
\end{equation}
up to a multiplicative constant depending on $\alpha$ and $\nu$. The thesis follows straightforwardly after a proper change of variable, $\eta\mapsto \frac{x}{2\sqrt{\nu-\delta_{\alpha,1}}}+\frac{1}{2}$, and normalization. Observe also that the thesis holds for all real values $\nu>1$.
\end{proof}
\section{Asymptotic behavior of $\beta$--triangles}\label{A-beta}
In this Appendix we prove Theorem \ref{ThBeta} on the asymptotic distribution of $\beta$--triangles. We restate the theorem here for the reader's convenience.
\begin{thbeta2}
The family of triangles \eqref{P_nu_beta_normalizado} with $\nu\in\mathds N$ and $\beta>0$ satisfies the property
\begin{equation}\label{EqThB2}
\frac{N}{2\sqrt{\nu-\chi(\beta)}}\hat p_{N,n,\beta}^{(\nu)} \asym{N\gg 1} P_{q_\beta(\nu)}(x),
\end{equation}
where we have introduced the function
\begin{equation}\label{chi2}
\chi(\beta)\coloneqq 1+\delta_{\beta,1}-\max\left\{1,\frac{1}{\beta}\right\}=\begin{cases}0&\text{for $\beta>1$},\\1&\text{for $\beta=1$},\\1-\frac{1}{\beta}&\text{for $0<\beta<1$,} \end{cases}
\end{equation}
and the properly centered and rescaled variable
\begin{equation}\label{cov2}
x\coloneqq 2\sqrt{\nu-\chi(\beta)}\left(\frac{n}{N}-\frac{1}{2}\right).
\end{equation}
In Eq.~\eqref{EqThB2}, $P_{q_\beta(\nu)}(x)$ is a $q$--Gaussian with 
\begin{equation}
q_\beta(\nu)=1-\frac{1}{\nu-\chi(\beta)}=\begin{cases}1-\frac{1}{\nu}&\text{for $\beta>1$},\\1-\frac{1}{\nu-1}&\text{for $\beta=1$},\\1-\frac{\beta}{\beta\nu+1-\beta}&\text{for $0<\beta<1$.} \end{cases}\end{equation}
\end{thbeta2}
\begin{proof} The proof of the Theorem strictly follows the proof of Theorem \ref{ThAlpha} in \ref{A-alpha}, the only difference being the evaluation of the asymptotic behavior of the $\mu^{(\nu)}_{N,n,\beta}$ coefficient for large $N$ at fixed $\frac{n}{N}$. 

For $\beta=1$, we have simply $\mu^{(\nu)}_{N,n,1}=1$ so there is nothing to do. 

For $\beta\neq 1$ and $\nu=2,3,\dots$, denoting $N-n\eqqcolon N\eta$, we can write
\begin{equation}\label{muasym1}
\mu^{(\nu)}_{N,n,\beta}%\coloneqq\prod_{k=\nu}^{2\nu-1}b_k(\beta)\frac{\prod_{k=\nu}^{n+\nu-1}b_k(\beta)}{\prod_{k=N-n+\nu}^{N+2\nu-1}b_k(\beta)}
=\prod_{k=0}^{\nu-1}b_{k+\nu}(\beta)\frac{\prod_{k=0}^{N\eta-1}b_{k+\nu}(\beta)}{\prod_{k=0}^{N\eta-1+\nu}b_{k+\nu+N(1-\eta)}(\beta)}.
\end{equation}
Observe that the first product 
\begin{equation}
B_0\coloneqq \prod_{k=0}^{\nu-1}b_{k+\nu}(\beta)
\end{equation}
is only a global factor not depending on $N,n$. We need to perform our asymptotic analysis only on the fraction
\begin{equation}
\frac{\prod_{k=0}^{N\eta-1}b_{k+\nu}(\beta)}{\prod_{k=0}^{N\eta-1+\nu}b_{k+\nu+N(1-\eta)}(\beta)}.\label{frac}
\end{equation}
We distinguish the two cases, $0<\beta<1$ and $\beta>1$.

For $0<\beta<1$
\begin{equation}
\mu^{(\nu)}_{N,n,\beta}\asym{N\gg 1} B_- N^\frac{1}{\beta}\left[\eta(1-\eta)\right]^\frac{1}{\beta},\label{asmubm1}
\end{equation}
where $B_-$ is a certain constant. Indeed we have that the asymptotic behavior of the denominator of the fraction \eqref{frac} can be evaluated as
\begin{multline}
\frac{1}{\prod_{k=0}^{N\eta-1+\nu}b_{k+\nu+N(1-\eta)}(\beta)}\\
\sim \frac{1}{\beta^{N\eta+\nu}}\prod_{k=0}^{N\eta-1+\nu}\left(1-\frac{1}{\beta}\frac{1}{k+\nu+N(1-\eta)}\right)\\\textstyle
\\=\frac{\Gamma\left(N+2\nu-\frac{1}{\beta}\right)}{\beta^{N\eta+\nu}\Gamma(N+2\nu)}\frac{\Gamma\left(N(1-\eta)+\nu\right)}{\Gamma\left(N(1-\eta)+\nu-\frac{1}{\beta}\right)}\sim\frac{(1-\eta)^\frac{1}{\beta}}{\beta^{N\eta+\nu}}.
\end{multline}
Moreover, we have also
\begin{multline}
\prod_{k=0}^{N\eta-1}b_{k+\nu}(\beta)=\\=\beta^{N\eta}\prod_{k=0}^{N\eta-1}\frac{b_{k+\nu}(\beta)}{\beta}=\beta^{N\eta}\prod_{k=1}^\infty\frac{b_{k+\nu}(\beta)}{\beta b'_{k+\nu}(\beta)}.\end{multline}
In the previous expression we introduced
\begin{equation}
b'_{k+\nu}(\beta)\coloneqq\begin{cases}1&\text{for $1\leq k\leq N\eta-1$,}\\\frac{b_{k+\nu}(\beta)}{\beta}&\text{for $k\geq N\eta$.}\end{cases}
\end{equation}
In the large $N$ limit we have that 
\begin{multline}
\prod_{k=0}^{N\eta-1}b_{k+\nu}(\beta)=\beta^{N\eta}\prod_{k=1}^\infty\frac{b_{k+\nu}(\beta)}{\beta b'_{k+\nu}(\beta)}\\\asym{N\gg 1}\beta^{N\eta}\prod_{k=1}^\infty\frac{b_{k+\nu}(\beta)}{\beta b''_{k+\nu}(\beta)}=\beta^{N\eta} B_1(\beta)\prod_{k=1}^{N\eta-1}\left(1+\frac{1}{\beta(k+\nu)}\right)\\
=B_1(\beta)\beta^{N\eta}\frac{\Gamma(\nu)}{\Gamma\left(\nu+\frac{1}{\beta}\right)}\frac{\Gamma\left(N\eta+\nu+\frac{1}{\beta}\right)}{\Gamma\left(N\eta+\nu\right)}\\
\sim B_1(\beta)B_2(\beta)\beta^{N\eta} N^\frac{1}{\beta}\eta^\frac{1}{\beta},
\end{multline}
where we have introduced 
\begin{equation}
b''_{k+\nu}(\beta)\coloneqq\begin{cases}1&\text{for $1\leq k\leq N\eta-1$,}\\ 1+\frac{1}{\beta(k+\nu)}&\text{for $k\geq N\eta$,}\end{cases}
\end{equation}
%\\\sim B_1(\beta)\beta^{N\eta}\prod_{k=0}^{N\eta-1}\left(1+\frac{1}{\beta(k+\nu)}\right)=B_1(\beta)\beta^{N\eta}\frac{\Gamma(\nu)}{\Gamma\left(\nu+\frac{1}{\beta}\right)}\frac{\Gamma\left(N\eta+\nu+\frac{1}{\beta}\right)}{\Gamma\left(N\eta+\nu\right)}\\\sim B_1(\beta)B_2(\beta)\beta^{N\eta} N^\frac{1}{\beta}\eta^\frac{1}{\beta},\end{multline}
and the constants\footnote{Observe that the infinite product \eqref{B1} converges: indeed, this is is easily seen using the limit comparison test on the positive--terms series $-\ln B_1(\beta)$ with, e.g., the series $\sum_{k=1}^\infty\frac{1}{k^2}$.}
\begin{subequations}\label{Bconst}
\begin{align}\label{B1}
B_1(\beta)&\coloneqq \prod_{k=0}^{\infty}\frac{1+\frac{1}{k+\nu}-\frac{1-\beta}{1-\beta^{k+\nu}}}{\min\{\beta,1\}+\frac{1}{k+\nu}},\\ B_2(\beta)&\coloneqq \frac{\Gamma(\nu)}{\Gamma\left(\nu+\frac{1}{\beta}\right)e^\frac{1}{\beta}}.
\end{align}
\end{subequations}
Eq.~\eqref{asmubm1} follows directly identifying $B_-\equiv B_1(\beta)B_2(\beta)$.

For $\beta>1$ we have instead that
\begin{equation}
\mu^{(\nu)}_{N,n,\beta}\asym{N\gg 1} B_+ N\eta(1-\eta),
\label{asmubMaggiore1}\end{equation}
where $B_+$ is a certain constant depending on $\beta$. Indeed, in Eq.~\eqref{muasym1} we can write
\begin{multline}
\frac{\prod_{k=0}^{N\eta-1}b_{k+\nu}(\beta)}{\prod_{k=0}^{N\eta-1+\nu}b_{k+\nu+N(1-\eta)}(\beta)}\\\sim\prod_{k=0}^{N\eta-1}b_{k+\nu}(\beta)\prod_{k=0}^{N\eta-1+\nu}\left(1-\frac{1}{k+\nu+N(1-\eta)}\right)\\
=\frac{N(1-\eta)+\nu-1}{N+2\nu-1}\prod_{k=0}^{N\eta-1}b_{k+\nu}(\beta)\sim(1-\eta)\prod_{k=0}^{N\eta-1}b_{k+\nu}(\beta).
\end{multline}
The remaining product can be written similarly as
\begin{equation}
\prod_{k=0}^{N\eta-1}b_{k+\nu}(\beta)%=\frac{\prod_{k=0}^\infty\left(1+\frac{1}{k+\nu}-\frac{1-\beta}{1-\beta^{k+\nu}}\right)}{\prod_{k=[N\eta]}^\infty\left(1+\frac{1}{k+\nu}-\frac{1-\beta}{1-\beta^{k+\nu}}\right)}
\sim B_1(\beta)\prod_{k=0}^{N\eta-1}\left(1+\frac{1}{k+\nu}\right)\sim B_1(\beta)B_2(1) N\eta,
\end{equation}
where we used the definitions \eqref{Bconst}. Eq.~\eqref{asmubMaggiore1} follows directly imposing $B_+\equiv B_1(\beta)B_2(1)$.

Summarizing, for $\beta>0$, $\beta\neq 1$, up to a global multiplicative constant,
\begin{equation}
\mu^{(\nu)}_{N,n,\beta}\asym{N\gg 1}\left[N\eta(1-\eta)\right]^{\max\left\{1,\frac{1}{\beta}\right\}-\delta_{\beta,1}}.
\end{equation}
Defining the function $\chi(\beta)$ as in \eqref{chi2}, we have
\begin{equation}
N^{\chi(\beta)}\binom{N}{n}r^{(\nu)}_{N,n,\beta}\asym{N\gg 1}\left[\eta(1-\eta)\right]^{\nu-\chi(\beta)}.
\end{equation}
Introducing now the variable
\begin{equation}
x\coloneqq 2\sqrt{\nu-\chi(\beta)}\left(\eta-\frac{1}{2}\right)
\end{equation}
and properly normalizing we obtain the thesis.
\end{proof}
\bibliographystyle{elsarticle-harv}
\bibliography{Biblio.bib}
\end{document}